\documentclass[preprint,12pt]{elsarticle} 
\journal{}

\usepackage[utf8]{inputenc} 
\usepackage[T1]{fontenc}    
\usepackage{hyperref}       
\usepackage{url}            
\usepackage{booktabs}       
\usepackage{amsfonts}       
\usepackage{nicefrac}       
\usepackage{microtype}      
\usepackage{lipsum}
\usepackage{graphicx}       
\graphicspath{{media/}}     

\usepackage{amsmath}
\usepackage{subcaption}
\usepackage{tabularx}
\usepackage{array}
\usepackage{gensymb}
\usepackage[numbers]{natbib}

\begin{document}

\begin{frontmatter}
  
\title{Data-Driven Quantification of Battery Degradation Modes via Critical Features from Charging}

\author[1]{Yuanhao Cheng}
\ead{e1192594@u.nus.edu}
\author[1]{Hanyu Bai}
\ead{hanyuhanyu@u.nus.edu}
\author[1]{Yichen Liang}
\ead{yichen.liang@u.nus.edu}
\author[2]{Xiaofan Cui}
\ead{cuixf@seas.ucla.edu}
\author[3]{Weiran Jiang}
\ead{wjiang@farasis.com}
\author[1]{Ziyou Song\corref{cor1}}
\ead{ziyou@nus.edu.sg}
\affiliation[1]{organization={Department of Mechanical Engineering},
            addressline={National University of Singapore}, 
            city={Singapore},
            postcode={117575}, 
            country={Singapore}}

\affiliation[2]{organization={Department of Electrical and Computer Engineering},
            addressline={University of California, Los Angeles}, 
            city={Los Angeles},
            postcode={CA 90095}, 
            country={USA}}

\affiliation[3]{organization={Farasis Energy USA, Inc.},
            city={Hayward},
            postcode={CA 94545}, 
            country={USA}}

\cortext[cor1]{Corresponding author}



\begin{abstract}
Battery degradation modes influence the aging behavior of Li-ion batteries, leading to accelerated capacity loss and potential safety issues. Quantifying these aging mechanisms poses challenges for both online and offline diagnostics in charging station applications. Data-driven algorithms have emerged as effective tools for addressing state-of-health issues by learning hard-to-model electrochemical properties from data. This paper presents a data-driven method for quantifying battery degradation modes. Ninety-one statistical features are extracted from the incremental capacity curve derived from 1/3C charging data. These features are then screened based on dispersion, contribution, and correlation. Subsequently, machine learning models, including four baseline algorithms and a feedforward neural network, are used to estimate the degradation modes. Experimental validation indicates that the feedforward neural network outperforms the others, achieving a root mean square error of around 10\% across all three degradation modes (i.e., loss of lithium inventory, loss of active material on the positive electrode, and loss of active material on the negative electrode). The findings in this paper demonstrate the potential of machine learning for diagnosing battery degradation modes in charging station scenarios.
\end{abstract}

\begin{graphicalabstract}
\includegraphics[width=\textwidth]{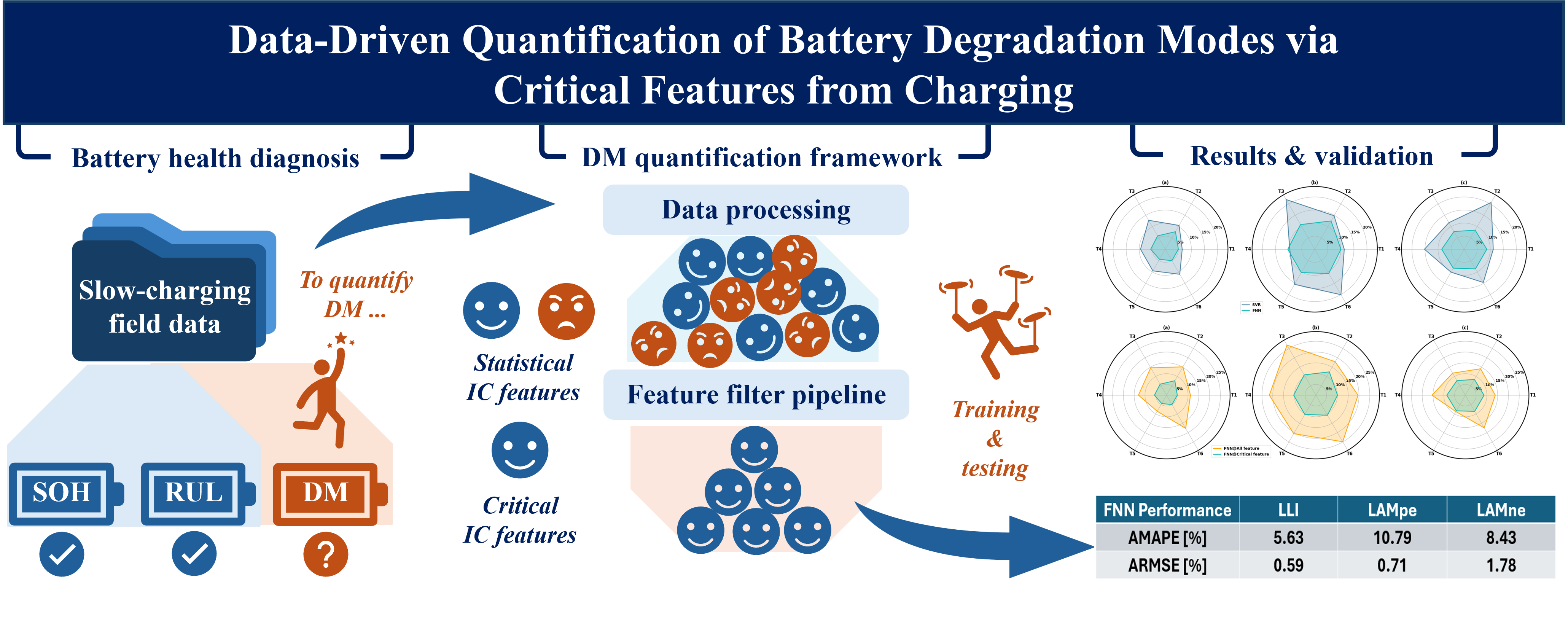}
\end{graphicalabstract}


\end{frontmatter}

\section*{Abbreviations}
\begin{tabular}{ll}
AF-FNN & All Feature Feedforward Neural Network \\
AMAPE & Average of Mean Absolute Percentage Error \\
ARMSE & Average of Root Mean Square Error \\
CCD & Constant-Current-Discharge \\
CF-FNN & Critical Feature Feedforward Neural Network \\
DCD & Drive-Current-Discharge \\
DM & Degradation Mode \\
DV & Differential Voltage \\
DVA & Differential Voltage Analysis \\
EFC & Equivalent Full Cycle \\
ENR & Elastic Net Regression \\
Expt4 & Experiment No.4 \\
Expt5 & Experiment No.5 \\
FNN & Feedforward Neural Network \\
ICA & Incremental Capacity Analysis \\
IC & Incremental Capacity \\
ICD & Differential IC Curve \\
LAM & Loss of Active Material \\
LAMne & Loss of Active Material in negative electrode \\
LAMpe & Loss of Active Material in positive electrode \\
LIB & Lithium-ion Battery \\
LLI & Loss of Lithium Inventory \\
MAPE & Mean Absolute Percentage Error \\
MLR & Multiple Linear Regression \\
OCV & Open Circuit Voltage \\
OCP & Open Circuit Potential \\
POCV & Pseudo Open Circuit Voltage \\
POCP & Pseudo Open Circuit Potential \\
RPT & Reference Performance Test \\
RMSE & Root Mean Square Error \\
RUL & Remaining Useful Life \\
SGPR & Sparse Gaussian Progress Regression \\
SOC & State of Charge \\
SOH & State of Health \\
SVR & Support Vector Regression \\
\end{tabular}

\section{Introduction}
The rapid development of lithium-ion batteries (LIBs) has reshaped the global energy resources market in recent decades. A wide range of application scenarios, from electric vehicles (EVs) to grid energy storage systems, benefit from various types of LIBs to achieve specific demands like high power output or large energy capacity \cite{nyamathulla2024review}. However, the performance of LIBs diminishes under varying working conditions over time. The state of health (SOH) is a universally accepted metric used to assess the degree of battery aging. Within the framework of SOH measurement, various degradation modes (DMs) critically influence the battery's health state and signify the tendency towards the battery's future degradation \cite{Dubarry2011evaluation}. Extraction of battery DM information could be critical in substantial applications \cite{Lucu2018}.

Numerous DMs have been extensively documented and analyzed in prior studies. To capture material fading within a cell, DMs are categorized into three main types: loss of lithium inventory (LLI), loss of active material on the positive electrode (LAMpe), and loss of active material on the negative electrode (LAMne) \cite{Birkl2016}. Within the past few years, there have been several studies focusing on DM analysis. Incremental capacity analysis (ICA) and differential voltage analysis (DVA) have been demonstrated as promising tools for characterizing DMs across various electrode chemistries \cite{Dubarry2022}. Dubarry et al. developed a diagnostic and prognostic model for battery degradation simulation in 2012 \cite{Dubarry2012}. Advanced data-driven methodologies have been employed to analyze synthetic degradation data \cite{Costa2024ICFormer}. Yet most of these studies predominantly rely on simulated pseudo open circuit potential/voltage (POCP/POCV) data. In reality, Obtaining OCV data is time-consuming \cite{Xiong2017} although several methods have been developed to accelerate the OCV extraction process \cite{Tong2015, Weng2014, Liu2021}. Furthermore, field data is easier to obtain than OCV data in real-world applications, yet little research replaces OCV data with field data to analyze DMs online.

The operating conditions in real-world applications significantly impact the detection of DMs. \cite{Li2024}. The discharge process is inherently dynamic, determined by user behavior, operational protocols, and time constraints. This variability presents challenges for conventional ICA and DVA, which rely on stable, quasi-open-circuit current flows. The slow charging process with a single-stage current, on the other hand, typically follows a stable and constant current charging procedure \cite{Khalid2019}, making it a potentially reliable signal for analyzing DMs. EVs charge protocol has been classified into different levels by the Society of Automatic Engineers \cite{SAEJ1772_2010}. Level-I and Level-II charging, which restrict charging times to 4-12 hours and 2-6 hours respectively, are classified as semi-fast charging. Conversely, Level-III charging is classified as fast charging, offering charging times ranging from 1/5 to 1 hour \cite{Su2012}. Real-world stationary charging rates, which are higher than POCV current flows, present distinct challenges for applying ICA to actual charging station data. These challenges include: (1) Are traditional IC features, such as specific IC peaks, still effective for analyzing DMs under real conditions? (2) What types of features should be extracted from higher C-rate IC curves? (3) How can the IC curve be effectively analyzed under non-POCV conditions?

Most existing IC curve features for DM analysis are related to IC peaks \cite{Dubarry2022}. Peak evolution, including position offset, value decline, and area shrinking, can correspond to degradation occurrence. Quantification of these peak behaviors presents degradation proportions of different battery components. Relevant studies have been done on a large variety of battery chemistry combinations, such as graphite intercalation compound/layered oxide-based batteries, Spinel-based batteries, Iron phosphate-based batteries, and Lithium titanate-based batteries \cite{Carter2021, Dubarry2020, Dubarry2011, Dubarry2009, Baure2020}. IC peak-based feature analysis begins with a specified degradation map, determined by the battery chemistry under investigation. These degradation maps are generated as compilations of synthetic curves representing POCV variations associated with single DM \cite{Dubarry2012}. When analyzing DMs under non-POCV scenarios, the standard degradation map potentially leads to inaccurate estimation results because the C-rate impacts peak locations. Some peaks may overlap or shift in different paths compared to the degradation map, and even disappear under charge station protocol C-rate or deep degradation \cite{Seo2022, WU2024234670, SCHMITT2023106517}.

The use of a statistical feature set may offer a more reliable approach for analyzing battery degradation. Related features have been applied to the battery SOH and remaining useful life (RUL) estimation. Features based on current, voltage, and temperature curves and their integral curves could be set as health indicators for battery energy storage systems \cite{Srivastava2024}. Aging features are extracted from these data sources and associated with statistics principles, such as values of maximum, minimum, mean, and variance \cite{Zhao2024}. IC curves under different temperature ranges and differential IC curves (ICD) could be informatics for feature excavation \cite{Yao2024, Wen2022}. A proper feature set after filter pipeline \cite{Wang2023} could be generated for a data-driven model including deep learning models to train for SOH or RUL information prediction \cite{Jiang2023}, such as knee point prediction \cite{Du2024}. The partial charging curve contains practical features as well, expanding the potential of utilizing statistical features as battery health monitor input\cite{Li2024}.

Therefore, this paper proposes a quantification framework for DM diagnosis using critical IC features obtained from a real-world charging profile, and the aging mechanism estimation performance of several data-driven algorithms is assessed. The main contributions of this paper are summarized in three points, including:

\begin{enumerate}
\item Extend ICA data resource for DM diagnosis from POCV level to 1/3C real-world charging profile. Different cycle profiles with different IC shifting orientations are investigated.
\item Several data-driven methods using statistical features calculated from the IC curve and its derivations are tested and compared for DM quantification.
\item The substitutability of critical statistical IC features to IC peaks for DM quantification is discussed.
\end{enumerate}

The remaining part of this paper is organized as follows: Section 2 introduces the experimental information of the work, including an overview of the framework, dataset description, data processing, and feature extraction. Afterward, several methodologies, including multi-step feature filter pipeline, baseline algorithms, and feedforward neural network (FNN) are demonstrated in Section 3. Section 4 focuses on the results and their discussion. The conclusion is presented in in Section 5.

\section{Dataset Insights}
\subsection{Overview}
\begin{figure}[ht]
    \centering
    \includegraphics[width=\textwidth]{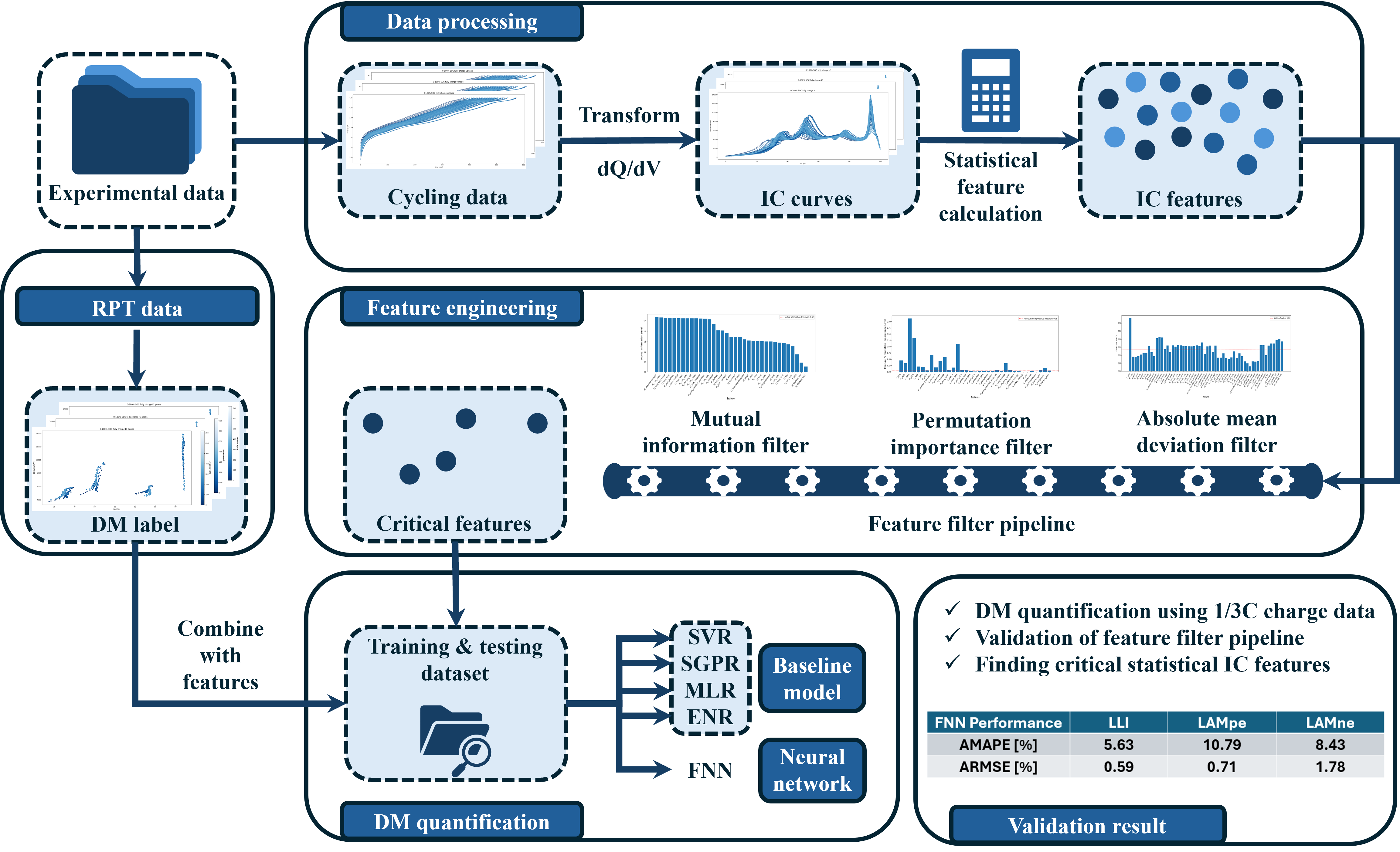}
    \caption{Framework of the proposed data-driven DM quantification}
    \label{Fig 1}
\end{figure}

The structure of this paper is shown in Fig.~\ref{Fig 1}. First, the original database selected from \cite{Kirkaldy2024} is processed, which contains two parts, the reference performance test (RPT) datasets and the aging datasets. The RPT datasets contain 1/5C and 1/20C charge-discharge cycles and the galvanostatic intermittent titration technique test. The aging datasets contain charging and discharging signals based on designed protocols. Degradation labels, including LLI, LAMpe, and LAMne values, are extracted from RPT datasets, and aging datasets provide IC curves from cycling data. More data description and processing details are described in Sections 2.2 and 2.3. Section 2.4 shows the extraction process of statistical IC feature sets, which are stored in the feature library. Critical features are chosen based on dispersion and correlation because not all features in the feature library would contribute to DM quantification. In this paper, "highly dynamic/active" means a high level of dispersion, and "highly contributed" means a strong correlation. A feature filter pipeline that consists of three different filter algorithms is used to select critical features. The first filter is an absolute mean deviation filter, which is used to verify the highly dynamic features. These features then pass through the other two feature filters, namely the permutation importance filter and the mutual information filter, to reserve the highly contributed features. Section 3.1 introduces the principles and application details of the feature filter pipeline. Critical features are selected by the feature filter pipeline and used as the inputs of machine learning models to estimate DMs. The performance of four baseline regression models, support vector regression (SVR), sparse Gaussian progress regression (SGPR), multiple linear regression (MLR), and elastic net regression (ENR), is compared with the FNN.

\subsection{Dataset Description}
In the original dataset, forty commercial 21700 lithium-ion cells with NMC811 as cathode and C/SiO\textsubscript{x} as anode were tested in five experiments with different setups, which cover operation conditions with four different state-of-charge (SOC) windows, 0-30\%, 70–85\%, 85–100\%, and 0–100\%, respectively, and three different temperatures [10, 25, 40]\degree C for each SOC window. The aging set is defined as a batch of cycles with a total throughput equal to seventy-eight equivalent full cycles (EFCs). DM values are extracted by RPT operated before and after every aging set. Each RPT contains a 1/10C fully charge-discharge cycle to calculate DM values \cite{Kirkaldy2022}. Experiment No.4 (Expt4) and Experiment No.5 (Expt5) are operated under the [0–100]\% SOC window with the same 1/3C charging process, which is assumed to satisfy the semi-fast charge standard described in Section 1. The difference between these two experiments is the discharge protocol, where Expt4 is under drive-cycle-discharge (DCD) based on the Worldwide Harmonized Light Vehicle Test Procedure \cite{Hales2021}, and Expt5 is under constant-current-discharge (CCD). In this work, the DCD and the CCD datasets are selected for DM quantification to investigate the characteristics of statistical features extracted from the [0–100]\% SOC window. Each dataset contains eight cells, which are allocated evenly under different temperature windows, meaning three cells for 10\degree C, two cells for 25\degree C, and three cells for 40\degree C. These cells are summarized in Table~\ref{Table 1}.

\subsection{Data Processing}
\begin{figure}
    \centering
    \includegraphics[width=1\textwidth]{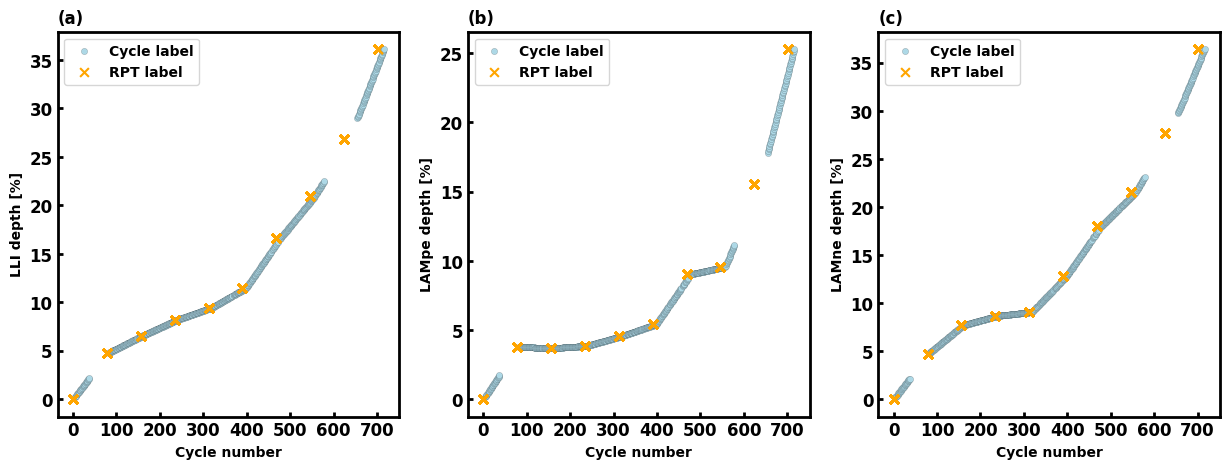}
    \caption{DM labels of cell CCD\textunderscore4. (a) LLI, (b) LAMpe, (c) LAMne. The cross marks in orange are real values from the RPT data, and the dots in blue are the interpolation results.}
    \label{Fig 2}
\end{figure}

In the original dataset, DMs are calculated only from RPTs. Therefore, to achieve the objective of DM quantification, cycle data in aging sets are labeled with virtual DM values based on real values from RPT data.
Assuming that the DM values grow monotonically through cycling. Fig.~\ref{Fig 2} shows an example of degradation labels from cell CCD\textunderscore4. The label lists have several gaps because the abnormal cycles with temperature control failure or electric signal record errors are removed. 

\begin{figure}
    \centering
    \includegraphics[width=0.7\textwidth]{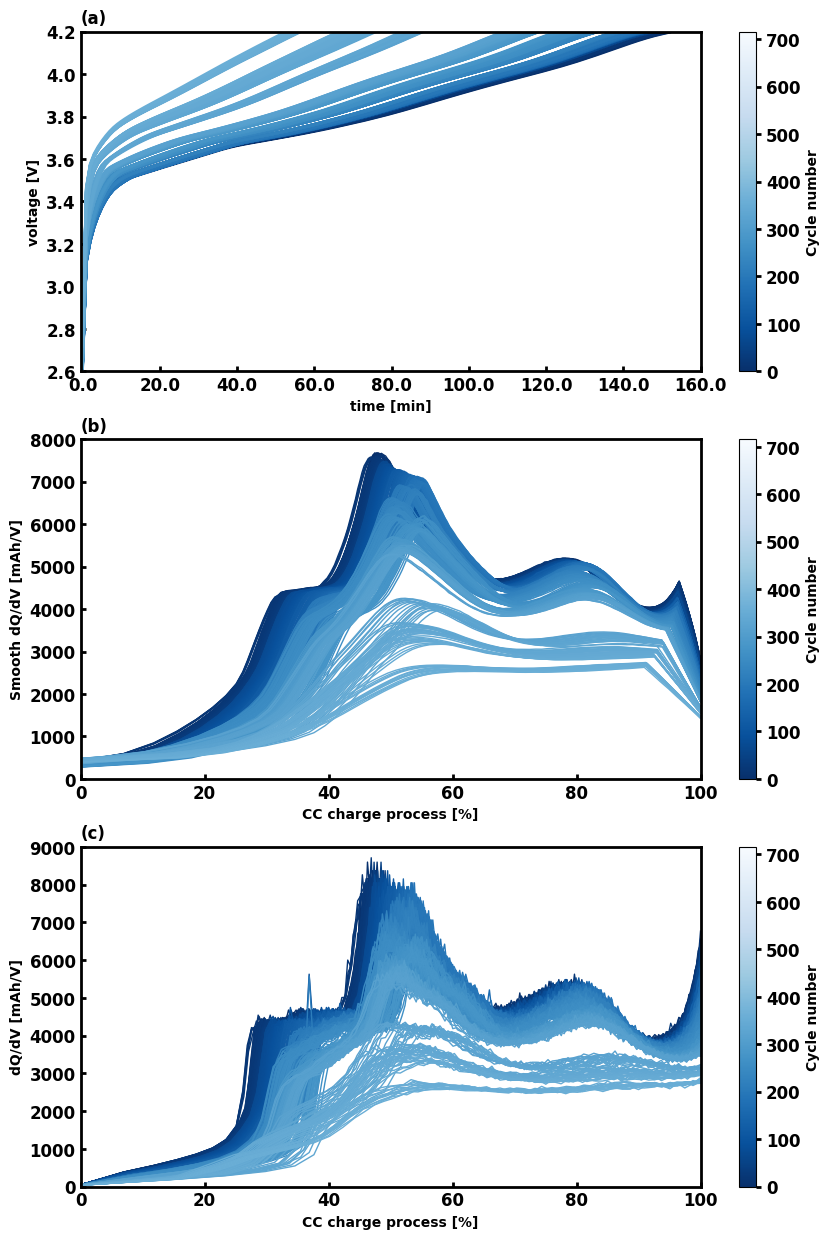}
    \caption{Information schematic of cell CCD\textunderscore4. (a) voltage, (b) smoothed IC, (c) original IC.}
    \label{Fig 3}
\end{figure}

Feature extraction is commonly done for time series data to extract electrochemistry-related features, specifically for current, voltage, temperature, and IC data \cite{Wang2023}. Fig.~\ref{Fig 3} shows the voltage and the IC curves of cell CCD\textunderscore4 as an example. The deep blue curve represents fresh cell data, and the light blue curve represents aged cell data. Fig.~\ref{Fig 3}(a) demonstrates that with an identical charging current, the charging process reaches the upper voltage threshold faster during cell aging. The IC curve is produced by differentiating the voltage-versus-capacity curve into the dQ/dV-versus-voltage curve, which carries electrochemical information of phase transition in peaks and valleys.
In previous studies, DM estimation using IC data focused on tracking IC peak locations, with noise filtering to ensure precise extraction. In this study, IC peaks are not selected as features; instead, more statistical IC features are explored despite the noises in the data. Hence, the original IC curve is preferred for feature discovery. Fig.~\ref{Fig 3}(b) and Fig.~\ref{Fig 3}(c) compare IC curves with and without smoothing. The smooth IC curve shows a lower peak position at the last 10\% of the charging process and fails to demonstrate the disappearance of the final peak when the cell ages deeply. In the middle of the IC curve (around 30\% to 80\% of the charging process), peak locations overlap and disappear when severe degradation happens, making peak extraction difficult. Therefore, the peak and valley locations are not used as features, and the original IC curve is a more promising feature extraction resource than the smooth IC curve considering low SOH circumstances. 

\begin{table}
\centering
\scriptsize
\begin{tabular}{cccc}
    \toprule
    Test protocol & 10\degree C & 25\degree C & 40\degree C \\
    \midrule
    CCD & CCD\textunderscore1, CCD\textunderscore2, CCD\textunderscore3  & CCD\textunderscore4, CCD\textunderscore5 & CCD\textunderscore6, CCD\textunderscore7, CCD\textunderscore8 \\
    DCD & DCD\textunderscore1, DCD\textunderscore2, DCD\textunderscore3  & DCD\textunderscore4, DCD\textunderscore5 & DCD\textunderscore6, DCD\textunderscore7, DCD\textunderscore8     \\
    \bottomrule
\end{tabular}
\caption{Cells in CCD and DCD datasets}
\label{Table 1}
\end{table}

\begin{figure}
    \centering
    \includegraphics[width=1\textwidth]{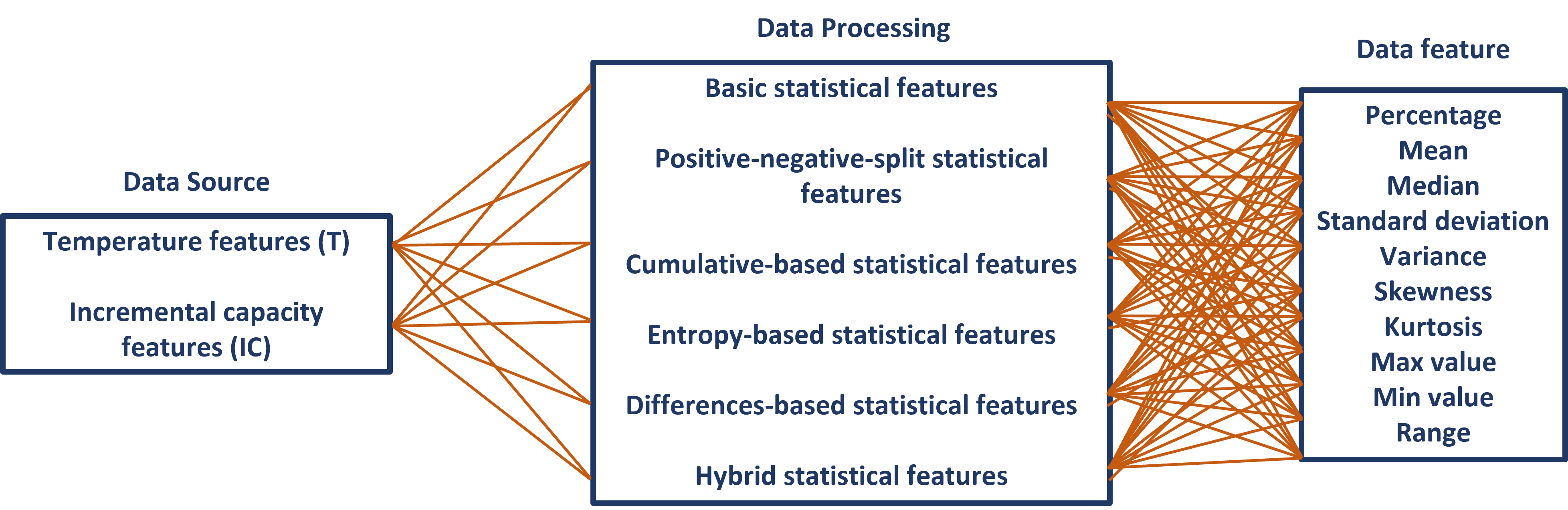}
    \caption{The feature library is categorized into three levels: data source, data processing, and data feature}
    \label{Fig 4}
\end{figure}

\subsection{Statistical Feature Extraction}
A feature library shown in Fig.~\ref{Fig 4} describes the generation of the statistical features. In the first level, features are categorized into a temperature feature set and an IC feature set considering data sources. In the second level, each set concludes six statistical subsets, which are the basic, positive-negative-split, cumulative-based, entropy-based, differences-based, and hybrid formation of data sources. These subsets describe the transformations applied to data resource curves. For example, IC's basic statistical feature subset loads features extracted from the IC curve. IC's positive-negative-split statistical feature subset carries features extracted from the positive and negative parts of the IC curve separately. IC's entropy-based statistical feature subset contains features extracted from the entropy of the IC curve. Note that the IC's hybrid statistical features are generated from throughput multiplied by other features. The third level contains specific parameters of features, including maximum, minimum, mean, median, percentage values, etc.. After calculations based on the three steps, ninety-one features are created in total.

\section{Methodology}
\subsection{Multi-step Feature Filter Pipeline}
In this study, the critical features satisfy the following criteria: (1) they have a high dispersion through battery aging, (2) they contribute to the machine learning estimation process efficiently, and (3) they are highly correlated to DMs. An adaptive multi-step feature filter pipeline with threshold control is proposed to screen critical features. The pipeline consists of three different filters with different functions. The first one is an absolute mean deviation filter, which extracts the features with a high dispersion coefficient through test cycles. The second filter is the permutation importance filter. It evaluates feature contribution to DM estimation generated by the extreme gradient boost (XGB), a decision tree-based algorithm. The final filter is the mutual information filter, which identifies features with a high correlation coefficient with estimation results. 

\subsubsection{Absolute Mean Deviation Filter}
In feature engineering, the absolute mean deviation value reflects the dispersion coefficient of a feature. A feature potentially shares an essential correlation with the estimation target if the feature's dispersion coefficient is high. On the contrary, small dispersion represents that the feature's values rarely change during battery aging and contain less useful information, which misleads model training and increases computation complexity. Thus, an absolute mean deviation filter is introduced to screen out these features from the feature library and reserve the features with high dispersion for later analysis. The determination of "high dispersion" is impacted by feature magnitudes, which cover a wide range of values. Therefore, before operating the feature filter pipeline, the feature library is normalized by using min-max normalization, which is shown in Eq~\ref{eq:minmax_norm}. $x_{\text{min}}$ and $x_{\text{max}}$ mean the minimum and maximum value of a feature, respectively. The normalized feature values are in a range between 0 and 1 based on percentile. Hence, the absolute mean deviation calculation would not be affected by magnitude differences. The absolute mean deviation equation is shown in Eq~\ref{eq:amd}. $\mu$ represents the mean value of a normalized feature, and $n$ is the number of values in the feature. Considering the following feature extraction demand, a relatively conservative screening ratio is decided to reserve the high dispersion features. After the first filter, sixty-three features out of ninety-one total features are submitted to the following filters for correlation coefficient analysis.

\begin{equation}
x_{\text{normalized}} = \frac{x - x_{\text{min}}}{x_{\text{max}} - x_{\text{min}}}
\label{eq:minmax_norm}
\end{equation}

\begin{equation}
x_{\text{AMD}}= \frac{1}{n} \sum_{i=1}^{n} |x_i - \mu|
\label{eq:amd}
\end{equation}

\subsubsection{Permutation Importance Filter}
To determine the eventual critical feature group after the high-dynamic feature pool is confirmed, the contribution to data-driven algorithms of the remaining characteristics should be evaluated from the importance and correlation perspectives. A practical principle is to disturb the correlation between the original prediction input and output and assess training results. Null importance, one of the methods based on the random search concept, has been proven to be a promising tool for achieving feature weight assessment for SOH estimation by Zhao et al. \cite{Zhao2024}. In their work, a gradient boosting machine called LightGBM is picked as a null importance calculator. Initially, the benchmark score is calculated by training with the feature set and the label set in the correct pair. Afterward, label elements are randomly disrupted, and the model is trained again. Here, the importance score is used to describe how a feature correlates with the label. Suppose the feature has a strong correlation with the label. In that case, its importance score drops after the model is trained with the disrupted label as the feature is unable to contribute to prediction. On the other hand, the importance score changes slightly or increases if the feature has a weak correlation with the label. This test procedure is repeated multiple times to ensure the scoring comparison result is robust. The final feature score is evaluated by importance split and gain corresponding to the label benchmark result and label disruption iteration results.

In this work, XGB is employed as the computational method. Compared to LightGBM, XGB consumes relatively more memory and time on massive data resources. However, the entire dataset contains only sixteen cells with around 10000 cycles, and each cycle contains a high-dynamic feature subset with a size of sixty-three. The dataset is relatively small, resulting in negligible differences in computational resource consumption between the two potential evaluation methods. Because XGB could achieve higher prediction accuracy by using pre-sorted and exact algorithms, XGB instead of LightGBM is selected. Compared to the null importance filter, the permutation importance filter directly evaluates feature significance by perturbing elements within specific subsets and measuring the subsequent impact on estimation accuracy. This procedure is repeated 100 times for each subset across all degradation label sets, using the mean score to assess each feature. Critical features significantly impact accuracy upon disturbance, resulting in high scores, while non-contributory features produce minimal changes and low scores. Unlike the null importance analysis, the permutation importance filter provides a straightforward assessment by requiring only one training session per DM label set and simplifying feature importance to a single-dimensional value per feature, making it an easily interpretable algorithm. Therefore, the permutation importance filter is applied.

In the previous absolute mean deviation filter, the standard deviation of IC curve IC\textunderscore standard\textunderscore deviation and the skewness of IC curve IC\textunderscore skewness earn high ranking. Further permutation importance assessment points out that IC\textunderscore standard\textunderscore deviation and IC\textunderscore skewness rank at the top, which continuously increases their possibility of contributing to DM quantification. On the contrary, fourteen out of sixteen of the features extracted from the IC cumulation curve are high-dynamic features, but only two of them earn higher scores than average in the permutation importance filter. Based on the above analysis, IC\textunderscore standard\textunderscore deviation and IC\textunderscore skewness are critical features for LLI estimation, while IC-cumulation-based features are still under observation.

\subsubsection{Mutual Information Filter}
In the previous section, the permutation importance filter selected the high-contributing features for machine learning from the high-dynamic features. To ensure the final feature set's reliability, correlation analysis should be conducted to confirm that high-correlating features are not missed. Several correlation analysis methods, including Pearson correlation analysis and Spearman correlation analysis, have been proven to be practical in literature. These two tools focus on digging into different sorts of correlating relationships. Pearson correlation analysis measures the linear correlation between two continuous variables, while Spearman measures the monotonic correlation between two ordinal or continuous variables.

In this study, the degradation label set refers to three dimensions, which are LLI, LAMpe, and LAMne. Feature correlations with each other and with DMs could be complicated, and DMs share correlations among themselves as well. Therefore, to tackle the three-dimensional challenge, mutual information analysis is introduced as the correlation filter. Developed from information theory, mutual information calculates the information quantity that a variable shares with another variable. It captures all sorts of correlations, including linear, non-linear, monotonic, and non-monotonic. The measurement is based on the probability distributions of the variables involved, where a higher mutual information value indicates a stronger correlation and more shared information between variables.

Unlike Pearson and Spearman, which take values from -1 to 1, the depth of mutual information degree is non-negative. The mutual information score is zero if the two variables share no information. The higher the scoring, the deeper the correlation between the analysis targets. This characteristic is similar to permutation importance. Therefore, the mutual information filter and permutation importance filter are evaluated under the identical rule, which is the feature score ranking and further algorithm performance. In the proposed method, since two second-step filters work in parallel, any feature passing either filter is considered a critical feature. The pool consists of fixed features and high-dynamic features that are either highly contributing or highly correlated. High-dynamic features that meet both criteria are also selected. For instance, in the IC-cumulation-based feature subset, while most elements fall below the permutation importance threshold, some rank highly based on mutual information. This includes metrics like mean, max value, 90th percentile, kurtosis, and median of the cumulative IC curve. However, IC\textunderscore standard\textunderscore deviation and IC\textunderscore skewness show lower mutual information shared with LLI while ranking highly in permutation importance. Therefore, they are determined as critical features. On the contrary, the throughput of IC curve IC\textunderscore thp earns low scores from both second-step filters, which indicates that it is a low-contributing and low-correlating feature. Thus, IC\textunderscore thp is not on the list of critical features.

Critical features for each DM are extracted using the feature filter pipeline. Once extracted, these critical features are combined into a unified set of critical features for all DMs. In total, twenty-one IC features are selected as the final critical features.

\subsection{Data-driven Estimation Models}
In the feature engineering process, the integral feature filter pipeline identified the critical features. The next step is to quantify battery DM using data-driven methods, including four baseline algorithms and an FNN. The baseline algorithms are SVR, SGPR, MLR, and ENR, respectively. SVR applies the principles of support vector machine to regression \cite{Wang2018}, aiming to fit errors within a specific threshold while being robust against outliers, making it a potentially efficient tool to model the relationship between features and degradation labels. SGPR optimizes Gaussian process methods by using a subset of data points, termed inducing points, to reduce computational demands significantly and accelerate the calculation process. The MLR establishes relationships between multiple independent variables and a dependent variable, using a linear equation fitted to the data to predict outcomes. Both SGPR and MLR are fast, but MLR is simpler. ENR combines the L1 and L2 penalties from lasso and ridge methods and is ideal for handling correlated features and balancing variable selection with model complexity.

These regression methods are designed to estimate values in a single dimension, meaning each of the three degradation label subsets must be estimated separately. However, DMs interact in complex ways, leading to overlapping electrical signals. Different combinations of DM values can exhibit similar or even identical behavior on the IC curve, potentially confusing regression model training. Therefore, an FNN is designed to estimate all three DMs at the same time. The network’s output is three-dimensional, corresponding to the values of LLI, LAMpe, and LAMne. A loss function, defined by Eq~\ref{loss}, is designed to improve accuracy for each DM. The $w$ represents penalty weights applied on DMs, and 1, 4, and 2 are assigned for LLI, LAMpe, and LAMne, respectively. The network architecture consists of five embedding layers and five prediction layers connected through fully connected layers.

\begin{scriptsize}
\begin{equation}
\begin{split}
\text{Training loss} = RMSE_{\text{LLI}} \cdot w_{\text{LLI}} + 
RMSE_{\text{LAMpe}} \cdot w_{\text{LAMpe}} + 
RMSE_{\text{LAMne}} \cdot w_{\text{LAMne}}
\end{split}
\label{loss}
\end{equation}
\end{scriptsize}

\section{Results and Discussions}
After feature engineering for critical feature selection, the quantification of battery DM is obtained by data-driven methods. The results are demonstrated from the following perspectives: performance of the four baseline models is first evaluated and ranked; the baseline model with the best performance is chosen as the representative regression model to compare with the FNN model; then, another test of performance with and without feature engineering is set to evaluate the efficiency of the feature filter pipeline and validate the critical features. Two error metrics, RMSE and mean absolute percentage error (MAPE), are used to determine estimation accuracy.

\begin{figure}
    \centering
    \includegraphics[width=\textwidth]{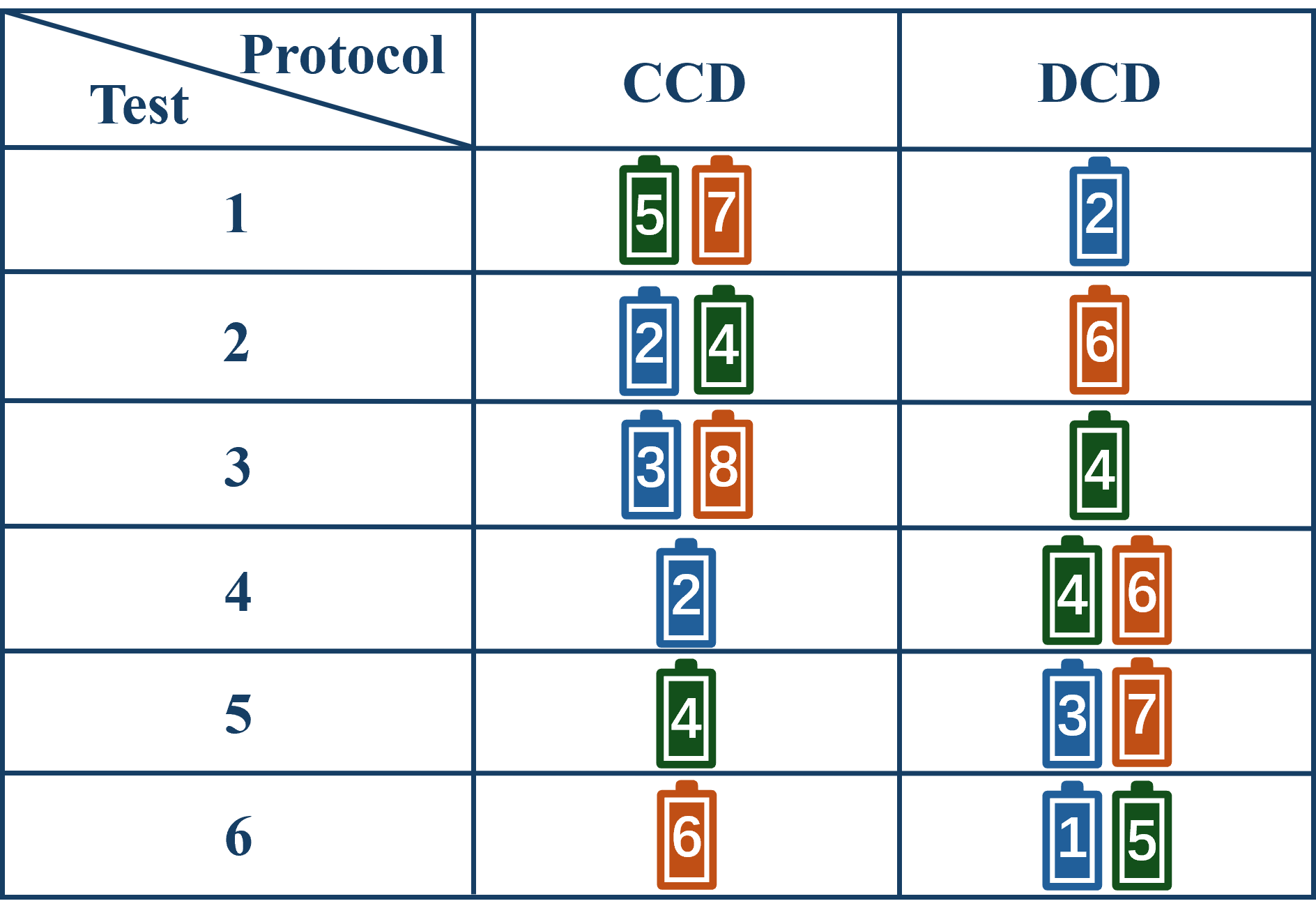}
    \caption{Summary of the test cells. The blue, green, and red cells represent 10\degree C, 25\degree C, and 40\degree C, respectively. Test cells are selected from both protocols and all temperature conditions.}
    \label{Fig 5}
\end{figure}

Due to the limitation of data, a statistical performance test series shown in Fig.~\ref{Fig 5} is designed to assess all the models applied in this work\cite{Cui2024, Cui20242}. The test series includes six tests with three cells per test. Both protocols, all three temperature conditions, and most cells are contained to ensure comprehensive test coverage. The average of MAPE (AMAPE), the average of RMSE (ARMSE), and the standard deviation of MAPE and RMSE from the six test combinations are calculated and given in Table~\ref{Table 2} and Table~\ref{Table 3}.

\begin{table}
  \centering
  \begin{tabular}{cccc}
    \toprule
    Model & LLI[\%] & LAMpe[\%] & LAMne[\%]\\
    \midrule
    SVR & 9.98/1.91 & 15.53/4.38  & 13.92/3.52\\
    SGPR & 11.51/1.36 & 17.17/3.11  & 17.67/4.44\\
    MLR & 12.73/1.87 & 18.87/2.15  & 20.40/4.62\\
    ENR & 17.28/3.50 & 23.46/2.33  & 23.03/4.98\\
    CF-FNN & 5.63/1.06 & 10.79/0.81  & 8.43/0.29\\
    AF-FNN & 13.40/2.95 & 21.90/3.06  & 13.60/2.75\\
    \bottomrule
  \end{tabular}
  \caption{Summary of model MAPE. For each result, the first number represents AMAPE and the second number represents the standard deviation of MAPE through the six tests. CF-FNN represents FNN using critical features, and AF-FNN represents FNN using all features.}
  \label{Table 2}
\end{table}

\begin{table}
  \centering
  \begin{tabular}{cccc}
    \toprule
    Model & LLI[\%] & LAMpe[\%] & LAMne[\%]\\
    \midrule
    SVR & 1.00/0.24 & 1.18/0.30  & 1.47/0.36\\
    SGPR & 1.01/0.26 & 1.09/0.34  & 1.66/0.36\\
    MLR & 1.11/0.12 & 1.16/0.29  & 2.01/0.29\\
    ENR & 1.68/0.30 & 1.13/0.13  & 2.06/0.42\\
    CF-FNN & 0.59/0.11 & 0.71/0.07  & 1.78/0.21\\
    AF-FNN & 2.18/1.27 & 3.54/3.14  & 4.18/3.48\\
    \bottomrule
  \end{tabular}
  \caption{Summary of model RMSE. For each result, the first number represents ARMSE and the second number represents the standard deviation of RMSE through the six tests. CF-FNN represents FNN using critical features, and AF-FNN represents FNN using all features.}
  \label{Table 3}
\end{table}

\subsection{DM Quantification Performance}
\subsubsection{Performance of Baseline Models}
A statistical significance t-test is applied to compare the performance of baseline models. In this work, the null hypothesis assumes that the performance of SVR is similar to the other baseline models, while the alternate hypothesis assumes that SVR performs better. The significance level is set as 5\%. The calculation of the t-value and degrees of freedom are given in Eq~\ref{t-value} and Eq~\ref{free}, where $\bar{x}_{\text{Model}}$ represents AMAPE of the model under comparison, $\bar{x}_{\text{SVR}}$ represents AMAPE of SVR, $s_{\text{Model}}$ represents the standard deviation of MAPE of the model under comparison, and $s_{\text{SVR}}$ represents the standard deviation of the MAPE of SVR.

The t-test results are shown in Table~\ref{Table 4}. Based on the significance level and the t-distribution table, the $t_{\text{threshold}}$ is 2.145. Since SVR is the benchmark in the hypotheses, its t-values are zeros. For LAMpe estimation, both SGPR and MLR have t-values below the $t_{\text{threshold}}$. However, for the total t-value, only SGPR meets the null hypothesis, indicating that SGPR and SVR are comparable. MLR and ENR have total t-values higher than the $t_{\text{threshold}}$, suggesting that SVR outperforms both MLR and ENR.

\begin{equation}
t = \frac{\bar{x}_{\text{Model}} - \bar{x}_{\text{SVR}}}{\sqrt{\frac{s_{\text{Model}}^2}{6} + \frac{s_{\text{SVR}}^2}{6}}}
\label{t-value}
\end{equation}

\begin{equation}
df = n + n - 2 = 6 + 6 - 2 = 10
\label{free}
\end{equation}

\begin{table}
  \centering
  \begin{tabular}{ccccc}
    \toprule
    Model & LLI & LAMpe & LAMne & Total\\
    \midrule
    SVR & 0 & 0 & 0 & 0\\
    SGPR & 1.598 & 0.748 & 1.621 & 1.322\\
    MLR & 2.520 & 1.677 & 2.733 & 2.310\\
    ENR & 4.485 & 3.915 & 3.659 & 4.020\\
    \bottomrule
  \end{tabular}
  \caption{T-test values of baseline models. The total t-test score is the mean value of each model's t-values on the three DMs.}
  \label{Table 4}
\end{table}

\begin{figure}
    \centering
    \includegraphics[width=\textwidth]{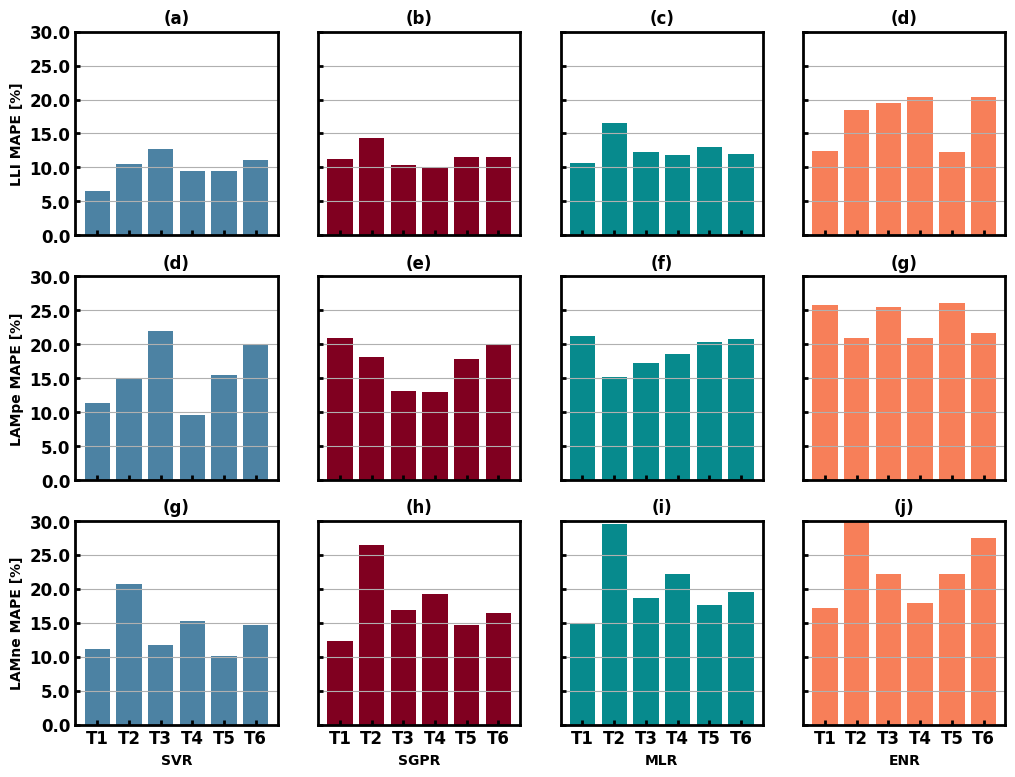}
    \caption{MAPE of Baseline models. Each bar represents the MAPE of a test. Columns from left to right are the MAPE of SVR, SGPR, MLR, and ENR, respectively. Rows from upside to downside are the MAPE of LLI, LAMpe, and LAMne, respectively.}
    \label{Fig 6}
\end{figure}

The MAPE distribution of the four regression models is assessed and shown in Fig.~\ref{Fig 6}. LLI can be better estimated than LAMpe and LAMne based on the lower average height of bars. The accuracy of SVR is more stable than that of the other baseline models, showing fewer results exceeding 20\% of MAPE. This may be attributed to the adaptability of SVR in handling data that encompasses both linear and non-linear components. The DMs increase relatively linearly before severe degradation occurs. After approaching a knee point, different DM may grow at different speeds depending on conditions, leading to various non-linear changes.

\begin{figure}
    \centering
    \includegraphics[width=1\textwidth]{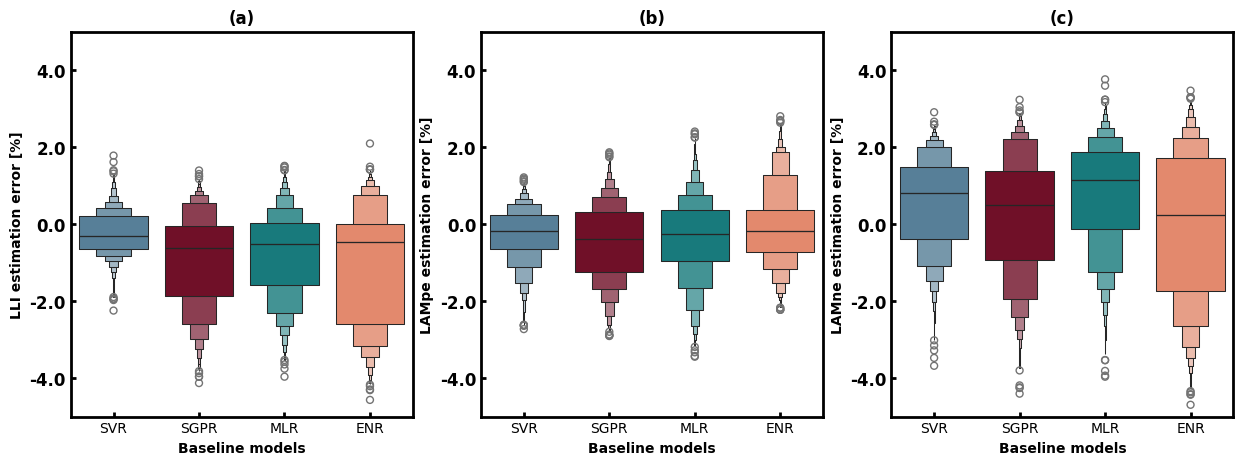}
    \caption{Baseline error boxplot of test 1. (a), (b), and (c) are the results of LLI, LAMpe, and LAMne, respectively. These enhanced box plots provide information on the distribution of data by plotting quantiles. Each error dataset is split into 15 boxes. The height of each box group represents the coverage of the estimation error space, and the area of each small box in a box group represents the error ratio under each interval.}
    \label{Fig 7}
\end{figure}

The LAMpe and LAMne estimation errors show greater variability compared to LLI estimation errors, particularly as batteries approach their knee points. The post-knee point accelerations, influenced by DM interactions, present a complex dynamic that is challenging for regression models. These models estimate each DM independently, without accounting for the beneficial insights gained from considering the interactions between DMs. As illustrated in Fig.~\ref{Fig 7}, which shows the distribution boxplot of baseline model errors in test 1, the ranges of LAMne quantification errors are larger than the other two DMs. This may be partly explained by the larger maximum value of LAMne than that of the other two DMs. On the contrary, the LAMpe boxplot covers a smaller area, which is partly correlated to the fact that LAMpe is smaller than LAMne during the battery lifetime. Hence, it can be explained that the MAPE of LAMpe quantification is the largest.

Based on the t-test result and the error distribution map, SVR performs the best among the baseline models, with AMAPE for LLI, LAMpe, and LAMne being 9.98\%, 15.53\%, and 13.92\%, and ARMSE being 1.00\%, 1.18\%, and 1.47\%, respectively.

\subsubsection{Performance of FNN}
\begin{figure}
    \centering
    \includegraphics[width=1\textwidth]{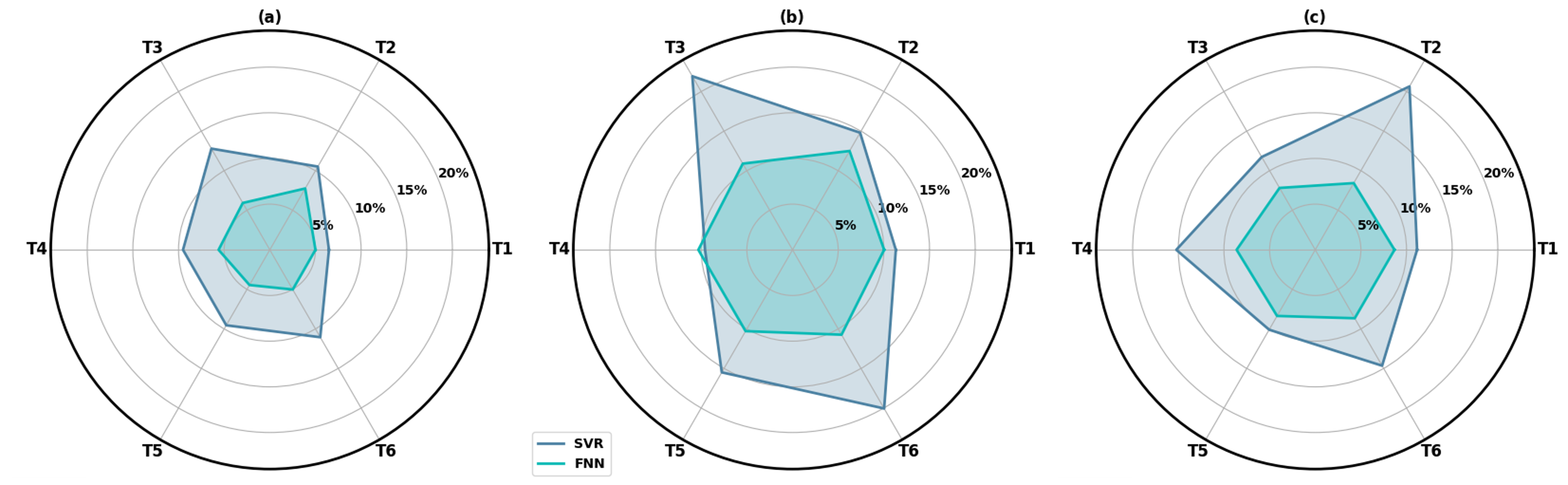}
    \caption{MAPE comparison between FNN and SVR. (a), (b), and (c) are the results of LLI, LAMpe, and LAMne, respectively. The blue area and the green area represent the SVR estimation MAPE and the FNN estimation MAPE, respectively. The size of the area is proportional to the MAPE value, larger areas correspond to higher MAPE.}
    \label{Fig 8}
\end{figure}

In the previous section, four regression models advantageous in single target estimation are applied to quantify DMs. These models are operated to estimate DMs one by one. For example, an SVR model is trained for LLI quantification. After the training and testing results are generated, another SVR model is trained for LAMpe or LAMne quantification. Each baseline model is trained three times to estimate the three labels separately. However, when a LIB is aging, DMs grow simultaneously and impact each other. Quantifying DMs one by one may neglect the interaction among battery interior electrochemical processes. To quantify DMs considering mechanism-to-mechanism interaction, an FNN is developed with the final output containing three objectives corresponding to LLI, LAMpe, and LAMne. The MAPE of FNN is compared with the best baseline model SVR, as shown in Fig.~\ref{Fig 8}. The MAPE of FNN is generally lower than that of SVR, as reflected in the smaller area shown in the radar maps. The structure of FNN, with multiple fully connected layers and more adjustable parameters than SVR, offers enhanced flexibility, making it well-suited for capturing complex patterns in a larger dataset compared to the one applied in this work. The AMAPE values of FNN are 5.63\%, 10.79\%, and 8.43\% for LLI, LAMpe, and LAMne, respectively, all of which are lower than those of SVR. This indicates that FNN is more robust than SVR. In this study, the dataset is limited by the number of cells, DM depths, and the range of conditions. Expanding the dataset could enhance the performance of FNN. Furthermore, FNN is more suited than SVR for this task, as it effectively handles the simultaneous quantification of the three DMs.

\subsection{Feature Filter Efficiency and Critical Feature Validation}
\begin{figure}
    \centering
    \includegraphics[width=1\textwidth]{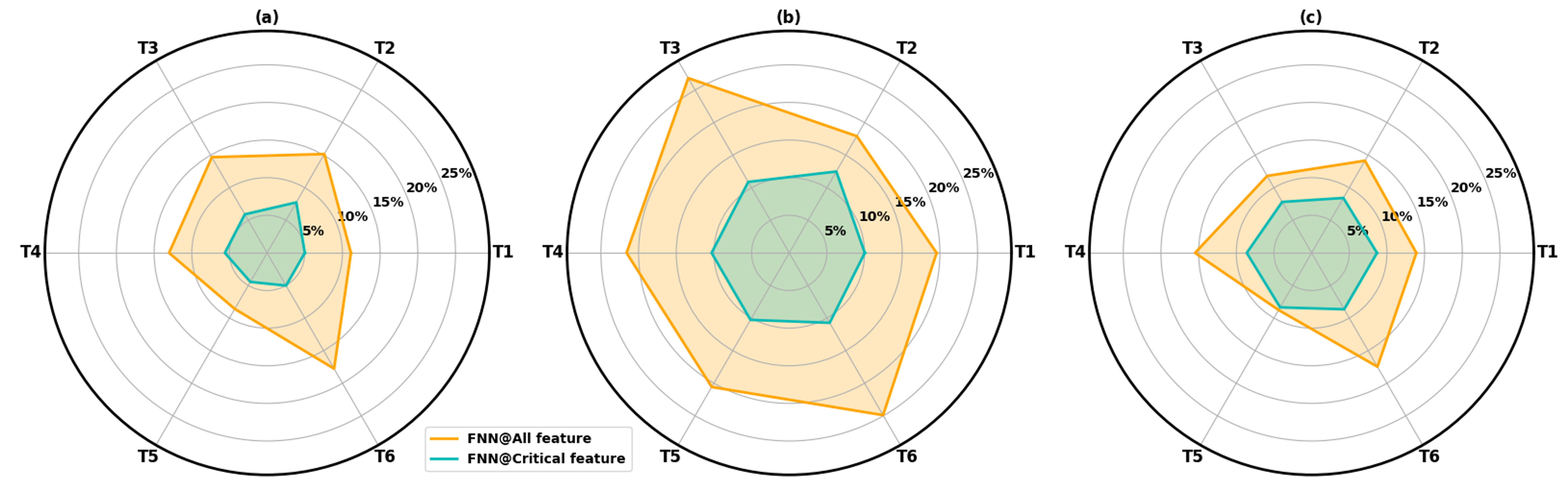}
    \caption{MAPE comparison between all feature FNN and critical feature FNN. (a), (b), and (c) are the results of LLI, LAMpe, and LAMne, respectively. The orange area and the green area represent the all-feature FNN estimation MAPE and the critical-feature FNN estimation MAPE, respectively. The size of the area is proportional to the MAPE value, larger areas correspond to higher MAPE.}
    \label{Fig 9}
\end{figure}

To evaluate the feature filter pipeline's efficiency, an FNN model using all features was trained for comparison. The MAPE comparison shown in Fig.~\ref{Fig 9} indicates that the feature filter improves estimation accuracy. The FNN with all features exhibits more dispersed estimations compared to the FNN using critical features, particularly in early degradation stages. Specifically, before LLI, LAMpe, and LAMne reach 5\%, 2\%, and 8\%, respectively, the FNN with all features generates significantly larger errors. After applying the feature filter, both MAPE and RMSE are reduced across all three DMs.

\begin{figure}
    \centering
    \includegraphics[width=0.7\textwidth]{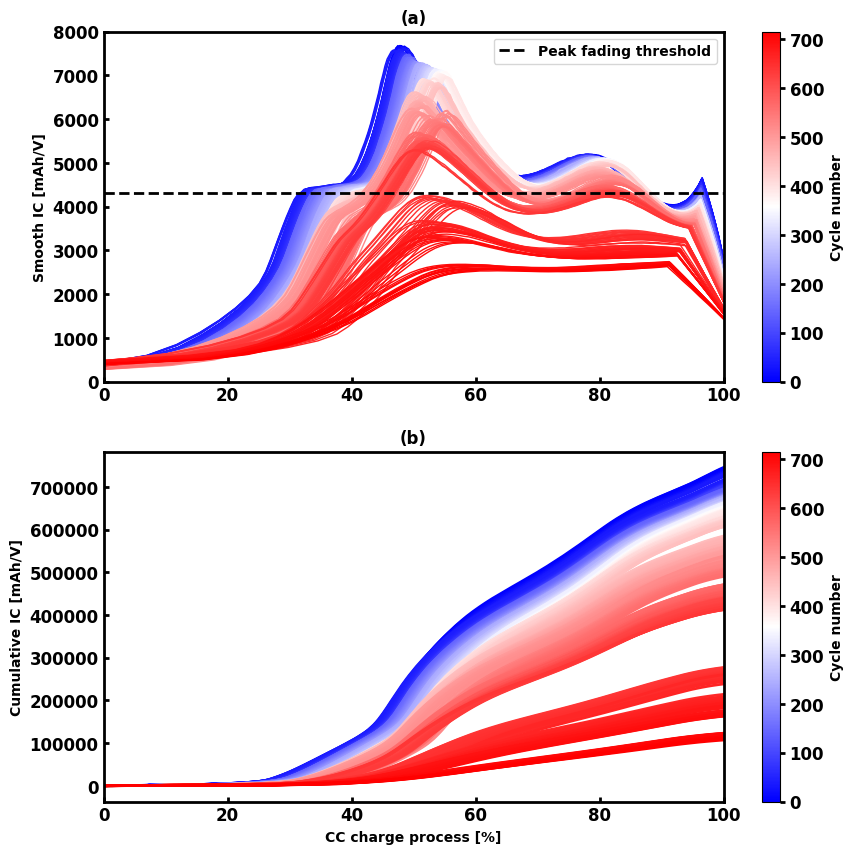}
    \caption{IC curve transformation formats: (a) smooth IC curves, (b) cumulative IC curves.}
    \label{Fig 10}
\end{figure}

The critical IC features, summarized in Table~\ref{Table 5}, can be classified into three categories: the IC-based features, the cumulative IC-based features, and the other features. Fig.~\ref{Fig 10} shows two of the IC transformation formats: the smooth IC curves and the cumulative IC curves. As illustrated in Table~\ref{Table 5}, most of the critical features are extracted from the IC curves and the cumulative IC curves, which illustrates that these two types of data may contain a more informative correlation with battery degradation than other IC transformation formats. Only one feature is extracted from the differential IC curves and two from throughput calculated by the IC curves.

In the IC-based critical feature category, six out of eight features are percentile features and situated near the IC peaks and valleys. This illustrates that IC percentile features can replace IC peaks and valleys in DM estimation. As demonstrated in Fig.~\ref{Fig 10}(a), the IC peaks shift and overlap when the cycle number increases. When the highest IC peak falls below 4300 mAh per volt, IC peaks in the 60\% to 90\% SOC area become indiscernible, potentially leading to failures in IC peak extraction. On the contrary, the IC percentile features correspond to certain locations on an entire IC curve. They are easy to obtain and will not disappear after severe degradation. In the cumulative IC-based critical feature category, the number of percentile features is three (70\%, 80\%, and 90\%), which is less than that in the IC-based critical feature category. The cumulative IC curve is similar to the voltage-versus-capacity curve in shape and characteristics. As shown in Fig.~\ref{Fig 10}(b), the percentile features in the lower SOC area change slightly while the ones in the higher SOC area change more aggressively. This can be explained by the significant capacity fade after a large number of cycles. In the meantime, several features, including the maximum value and the range of value, are sensitive to capacity fade as well. Therefore, these features are listed as critical features.

\begin{table}
  \centering
  \begin{tabular}{ccc}
    \toprule
    IC-based & Cumulative IC-based & Other\\
    \midrule
    30\% & 70\%  & IC\textunderscore diff\textunderscore 90\%\\
    40\% & 80\%  & IC\textunderscore thp\\
    50\% & 90\%  & IC\textunderscore thp*pos(mean)\\
    70\% & mean  & \\
    80\% & median  & \\
    90\% & std\textunderscore dev  & \\
    std\textunderscore dev & var  & \\
    var & kurtosis  & \\
     & max  & \\
     & range  & \\
    \bottomrule
  \end{tabular}
  \caption{Critical IC features}
  \label{Table 5}
\end{table}

\section{Conclusion}
This paper presents a data-driven approach to quantifying battery DMs using critical IC features, validated with experimental data. A feature filter pipeline is employed to identify the most relevant features. In the initial performance test, four regression models are trained and compared for DM quantification, followed by the development of an FNN model, which is compared to the best regression model results. The FNN model using critical IC features is further evaluated by training another FNN model with the full feature set. The results demonstrate that statistical features derived from the 1/3C IC curve provide sufficient information for accurate DM estimation. The proposed feature filter, based on information theory, enhances the accuracy of degradation quantification while reducing computational load. In this study, SVR is the best regression model, with AMAPE of LLI, LAMpe, and LAMne being 9.98\%, 15.53\%, and 13.92\%, respectively. While SVR performs well among the regression models, the multi-dimensional nature of the degradation targets makes the FNN a more efficient and appropriate choice. Compared to SVR, the AMAPE values of FNN are reduced to 5.63\%, 10.79\%, and 8.43\% for LLI, LAMpe, and LAMne, respectively. This study establishes a benchmark for battery DM quantification using data-driven methods and introduces critical IC features for further exploration.

\section*{CRediT authorship contribution statement}
\textbf{Yuanhao Cheng}: Writing – original draft, Software, Methodology, Conceptualization. 
\textbf{Hanyu Bai}: Writing – review \& editing, Methodology, Validation. 
\textbf{Yichen Liang}: Writing – review \& editing, Methodology, Validation. 
\textbf{Xiaofan Cui}: Writing – review \& editing, Supervision.
\textbf{Weiran Jiang}: Writing – review \& editing, Supervision.
\textbf{Ziyou Song}: Writing – review \& editing, Supervision.

\section*{Declaration of competing interest}
The authors declare that they have no known competing financial interests or personal relationships that could have appeared to
influence the work reported in this paper.

\newpage
\bibliographystyle{unsrt}  
\bibliography{references}  

\begin{thebibliography}{10}

\bibitem{nyamathulla2024review}
S.~Nyamathulla and C.~Dhanamjayulu.
\newblock A review of battery energy storage systems and advanced battery management system for different applications: Challenges and recommendations.
\newblock {\em Journal of Energy Storage}, 86:Article 111179, 2024.

\bibitem{Dubarry2011evaluation}
M.~Dubarry, C.~Truchot, B.~Y. Liaw, K.~Gering, S.~Sazhin, D.~Jamison, and C.~Michelbacher.
\newblock Evaluation of commercial li-ion cells based on composite electrode for plug-in evs.
\newblock {\em Journal of Power Sources}, 196:10336--10343, 2011.

\bibitem{Lucu2018}
M.~Lucu, E.~Martinez-Laserna, I.~Gandiaga, and H.~Camblong.
\newblock A critical review on self-adaptive li-ion battery ageing models.
\newblock {\em Journal of Power Sources}, 401:85--101, 2018.

\bibitem{Birkl2016}
C.~Birkl, M.~Roberts, E.~McTurk, P.~Bruce, and D.~Howey.
\newblock Degradation diagnostics for lithium ion cells.
\newblock {\em Journal of Power Sources}, 341:373--386, 2016.

\bibitem{Dubarry2022}
M.~Dubarry and D.~Beck.
\newblock Best practices for incremental capacity analysis.
\newblock {\em Frontiers in Energy Research}, 10:Article 1023555, 2022.

\bibitem{Dubarry2012}
M.~Dubarry, C.~Truchot, and B.~Y. Liaw.
\newblock Synthesize battery degradation modes via a diagnostic and prognostic model.
\newblock {\em Journal of Power Sources}, 219:204--216, 2012.

\bibitem{Costa2024ICFormer}
N.~Costa, D.~Anse{\'a}n, M.~Dubarry, and L.~S{\'a}nchez.
\newblock Icformer: A deep learning model for informed lithium-ion battery diagnosis and early knee detection.
\newblock {\em Journal of Power Sources}, 592:Article 233910, 2024.

\bibitem{Xiong2017}
Rui Xiong, Quanqing Yu, Le~Yi Wang, et~al.
\newblock A novel method to obtain the open circuit voltage for the state of charge of lithium ion batteries in electric vehicles by using h infinity filter.
\newblock {\em Applied Energy}, 207:346--353, 2017.

\bibitem{Tong2015}
Shijie Tong, Matthew~P. Klein, and Jae~Wan Park.
\newblock On-line optimization of battery open circuit voltage for improving state-of-charge and state-of-health estimation.
\newblock {\em Journal of Power Sources}, 2015.

\bibitem{Weng2014}
Chao Weng, Jiuchun Sun, and Huei Peng.
\newblock A unified open-circuit-voltage model of lithium-ion batteries for state-of-charge estimation and state-of-health monitoring.
\newblock {\em Journal of Power Sources}, 258:228--237, 2014.

\bibitem{Liu2021}
Y.~Liu, C.~T. Ke, L.~Y. Yang, H.~Liu, Y.~L. Chen, and J.~H. Yuan.
\newblock State of charge estimation of lithium-ion batteries using the open-circuit voltage at various ambient temperatures.
\newblock {\em Journal of Power Sources}, 482:228975, 2021.

\bibitem{Li2024}
X.~Li, D.~Yu, S.~B. Vilsen, and D.~I. Stroe.
\newblock Accuracy comparison and improvement for state of health estimation of lithium-ion battery based on random partial recharges and feature engineering.
\newblock {\em Journal of Energy Chemistry}, 92:591--604, 2024.

\bibitem{Khalid2019}
M.~R. Khalid, M.~S. Alam, A.~Sarwar, and M.~S.~J. Asghar.
\newblock A comprehensive review on electric vehicles charging infrastructures and their impacts on power-quality of the utility grid.
\newblock {\em eTransportation}, 1:Article 100006, 2019.

\bibitem{SAEJ1772_2010}
{SAE International}.
\newblock Sae electric vehicle and plug-in hybrid electric vehicle conductive charge coupler.
\newblock SAE Standards, Jan 2010.

\bibitem{Su2012}
W.~Su, H.~R. Eichi, W.~Zeng, and M.~Y. Chow.
\newblock A survey on the electrification of transportation in a smart grid environment.
\newblock {\em IEEE Transactions on Industrial Electronics}, 8:1--10, 2012.

\bibitem{Carter2021}
R.~Carter, T.~A. Kingston, R.~W. Atkinson, M.~Parmananda, M.~Dubarry, C.~Fear, P.~P. Mukherjee, and C.~T. Love.
\newblock Directionality of thermal gradients in lithium-ion batteries dictates diverging degradation modes.
\newblock {\em Cell Reports Physical Science}, 2:Article 100351, 2021.

\bibitem{Dubarry2020}
M.~Dubarry, G.~Baure, and A.~Devie.
\newblock Durability and reliability of ev batteries under electric utility grid operations: Path dependence of battery degradation.
\newblock {\em Energies}, 13:2494, 2020.

\bibitem{Dubarry2011}
M.~Dubarry, C.~Truchot, B.~Y. Liaw, K.~Gering, S.~Sazhin, D.~Jamison, and C.~Michelbacher.
\newblock Evaluation of commercial lithium-ion cells based on composite positive electrode for plug-in hybrid electric vehicle applications. part ii. degradation mechanism under 2c cycle aging.
\newblock {\em Journal of Power Sources}, 196:10355--10368, 2011.

\bibitem{Dubarry2009}
M.~Dubarry and B.~Y. Liaw.
\newblock Identify capacity fading mechanism in a commercial lifepo4 cell.
\newblock {\em Journal of Power Sources}, 194:541--549, 2009.

\bibitem{Baure2020}
G.~Baure and M.~Dubarry.
\newblock Battery durability and reliability under electric utility grid operations: 20-year forecast under different grid applications.
\newblock {\em Journal of Energy Storage}, 29:101391, 2020.

\bibitem{Seo2022}
G.~Seo, J.~Ha, M.~Kim, J.~Park, and J.~Lee.
\newblock Rapid determination of lithium-ion battery degradation: High c-rate lam and calculated limiting lli.
\newblock {\em Journal of Energy Chemistry}, 67:663--671, 2022.

\bibitem{WU2024234670}
Z.~Wu, Y.~Zhang, and H.~Wang.
\newblock Battery degradation diagnosis under normal usage without requiring regular calibration data.
\newblock {\em Journal of Power Sources}, 608:234670, 2024.

\bibitem{SCHMITT2023106517}
J.~Schmitt, M.~Rehm, A.~Karger, and A.~Jossen.
\newblock Capacity and degradation mode estimation for lithium-ion batteries based on partial charging curves at different current rates.
\newblock {\em Journal of Energy Storage}, 59:106517, 2023.

\bibitem{Srivastava2024}
A.~K. Srivastava and P.~Famouri.
\newblock Battery energy storage systems: A review of energy management systems and health metrics.
\newblock {\em Energies}, 17:1250, 2024.

\bibitem{Zhao2024}
B.~Zhao, W.~Zhang, Y.~Zhang, C.~Zhang, C.~Zhang, and J.~Zhang.
\newblock Research on the remaining useful life prediction method for lithium-ion batteries by fusion of feature engineering and deep learning.
\newblock {\em Applied Energy}, 358:122325, 2024.

\bibitem{Yao2024}
L.~Yao, J.~Wen, Y.~Xiao, C.~Zhang, Y.~Shen, G.~Cui, and D.~Xiao.
\newblock State of health estimation approach for li-ion batteries based on mechanism feature empowerment.
\newblock {\em Journal of Energy Storage}, 84:110965, 2024.

\bibitem{Wen2022}
J.~Wen, X.~Chen, X.~Li, and Y.~Li.
\newblock Soh prediction of lithium battery based on ic curve feature and bp neural network.
\newblock {\em Energy}, 261:125234, 2022.

\bibitem{Wang2023}
J.~Wang, C.~Zhang, L.~Zhang, X.~Su, W.~Zhang, X.~Li, and J.~Du.
\newblock A novel aging characteristics-based feature engineering for battery state of health estimation.
\newblock {\em Energy}, 273:127169, 2023.

\bibitem{Jiang2023}
Y.~Jiang, Y.~Chen, F.~Yang, and W.~Peng.
\newblock State of health estimation of lithium-ion battery with automatic feature extraction and self-attention learning mechanism.
\newblock {\em Journal of Power Sources}, 556:232466, 2023.

\bibitem{Du2024}
Jingcai Du and Caiping Zhang.
\newblock Aging abnormality detection of lithium-ion batteries combining feature engineering and deep learning.
\newblock {\em Energy}, 297:131276, 2024.

\bibitem{Kirkaldy2024}
Niall Kirkaldy, Mohammad~A. Samieian, Gregory~J. Offer, Monica Marinescu, and Yatish Patel.
\newblock Lithium-ion battery degradation: Comprehensive cycle ageing data and analysis for commercial 21700 cells.
\newblock {\em Journal of Power Sources}, May 2024.

\bibitem{Kirkaldy2022}
N.~Kirkaldy, M.~A. Samieian, G.~J. Offer, M.~Marinescu, and Y.~Patel.
\newblock Lithium-ion battery degradation: measuring rapid loss of active silicon in silicon-graphite composite electrodes.
\newblock {\em ACS Applied Energy Materials}, 5:13367--13376, 2022.

\bibitem{Hales2021}
A.~Hales, E.~Brouillet, Z.~Wang, B.~Edwards, M.~A. Samieian, J.~Kay, S.~Mores, D.~Auger, Y.~Patel, and G.~Offer.
\newblock Isothermal temperature control for battery testing and battery model parameterization.
\newblock {\em SAE International Journal of Electrified Vehicles}, 10:105--122, 2021.

\bibitem{Wang2018}
F.-K. Wang and T.~Mamo.
\newblock A hybrid model based on support vector regression and differential evolution for remaining useful lifetime prediction of lithium-ion batteries.
\newblock {\em Journal of Power Sources}, 401:49--54, 2018.

\bibitem{Cui2024}
X.~Cui, M.A. Khan, G.~Pozzato, S.~Singh, R.~Sharma, and S.~Onori.
\newblock Taking second-life batteries from exhausted to empowered using experiments, data analysis, and health estimation.
\newblock {\em Cell Reports Physical Science}, 5:101941, 2024.

\bibitem{Cui20242}
X.~Cui, M.~A. Khan, and S.~Onori.
\newblock Online adaptive data-driven state-of-health estimation for second-life batteries with bibo stability guarantees.
\newblock {\em arXiv preprint arXiv:2401.04734}, 2024.

\end{thebibliography}


\end{document}